\documentclass{elsart}
\usepackage{natbib}
\usepackage{graphicx}
\begin{document}
\runauthor{Christelle Roy}
\begin{frontmatter}
\title{STAR Results on Strangeness Production at RHIC energies}
\author{Christelle Roy}
\collab{STAR Collaboration}
\address{SUBATECH, Nantes, France}
%\thanks[list]{for the full author list see the contribution of
%M. Lamont, these proceedings}

\begin{abstract}
Strangeness measurement at RHIC energies constitutes one of the
favorite theme of the STAR Collaboration. Besides the fact that
strangeness enhancement has been proposed as a quark gluon plasma
signature, its production provides various and relevant information
about the collision evolution and especially on the hadronization
process. The investigation of short-lived strange particles as well as
their anti-particles and resonances allows an attempt of
characterization of the matter created in RHIC heavy ion collisions.
\end{abstract}
\end{frontmatter}

\section{Foreword}
RHIC physics program is devoted to the search of the quark gluon
plasma (QGP) keeping in mind that despite numerous heavy ion
experiments at CERN had the same goal, no unambiguous signal of its
formation could be highlighted~\cite{CERNreview}. From studies
performed at the AGS, SPS or RHIC facilities, one can outline that
before aspiring to a clear discovery of a transition phase sign, the
evolution of the fireball created in the collision has to be well
understood. Although it is trivial to say that the clue resides in the
characterization of hadrons because whether a QGP has been formed or
not, the collision story will end with a hadron production. The aim is 
to describe the various phases that the created system encounters,
within a characterization of the hadronization as complete as
possible. Let us briefly remind the scenario commonly adopted : the
pre-equilibrium phase, is related to the earlier times of the
collision, during which soft hard scatterings between partons
take place leading to high temperature and pressure. If the energy
density exceeds the critical density predicted by lattice
QCD~\cite{qcd}, a QGP may be formed. An expansion follows, 
system cools down until it encounters again the deconfinement
point. The hadronization process occurs, partonic degrees of
freedom are thus recombined in hadrons. These later still interact
(within elastic and inelastic scattering) until the chemical freeze-out :
at this point, the chemical composition of each hadron is fixed but
elastic scattering continue. Finally, the system encounters the
thermal (kinetic) freeze-out of the particle momenta~: particles stop
interacting and move freely, toward the detectors. It is still
unclear nowadays if the time between freeze-outs is short or not.

The present paper aims at answer to some fundamental questions such as
: what is the original environment for the particle creation, what is the nature of  
the production mechanisms, how much are the particles produced, what is the
degree of suddenness of the hadronization process ? The answers should
be found through strange particle measurements, also through their
anti-particles and their resonances, as described in the following
section.

\section{Story of the Eighties}
Some twenty years ago, Jan Rafelski and Berndt
M\"{u}ller~\cite{rafelski} proposed that the quantification of  
strangeness produced in the collision may allow to distinguish between
a hadronic gas and a quark gluon plasma. This QGP signature relies on
the fact that strangeness production may be enhanced if a QGP is
formed. Furthermore, the higher the strange content of the particle,
the higher the enhancement.

SPS experiments suited for strangeness measurements showed the
expected enhancement of the multi-strange particles~\cite{na57}, from
proton-nucleus (pA) to nucleus-nucleus (AA) collisions. However, it is
possible to explain this strangeness augmentation with purely hadronic
scenarios, such as, notably, the interpretation in terms of canonical
suppression~\cite{canonical}. A. Tounsi and collaborators pointed out
at the last Quark Matter conference~\cite{tounsi} that a transition
from canonical to the asymptotic grand canonical limit is consistent
with the recent NA49 results on $\Lambda$ enhancement at
$\sqrt{s_{NN}}$=40, 80, 158 GeV~\cite{na49}. In spite of the
numerous SPS results related to strangeness, further investigations
were highly required, and especially at RHIC, where the QGP is
expected to be formed during a longer time offering better conditions
for its discovery. 

STAR made one of its priority of this strangeness study, and what
will be discussed now, is the numerous interesting information
which have been provided for two years of data analyses on :
\begin{itemize}
\item anti-baryon over baryon ratios providing knowledge
ranging from the baryon density of the system to a hint on
production mechanisms. 
\item strange over non-strange particles allowing a quantitative
evaluation of strangeness production. 
\item ratios may also give access to an estimation of the
temperature and chemical potential at the chemical freeze-out. 
\item transverse mass or momentum distributions
informing on the partition between collective (flow velocity) and
thermal effects (temperature at kinetic freeze-out).
\item strange resonances serving as a probe for the duration of
the hadronization process,i.e. as a chronometer of the
time between the chemical and thermal freeze-outs.
\end{itemize}

\section{Hadron production}
\subsection{Original environment}
\begin{figure}
\begin{minipage}{0.47\linewidth}
\begin{center}
\hspace{-1.cm}
\includegraphics[height=\linewidth,width=\linewidth]{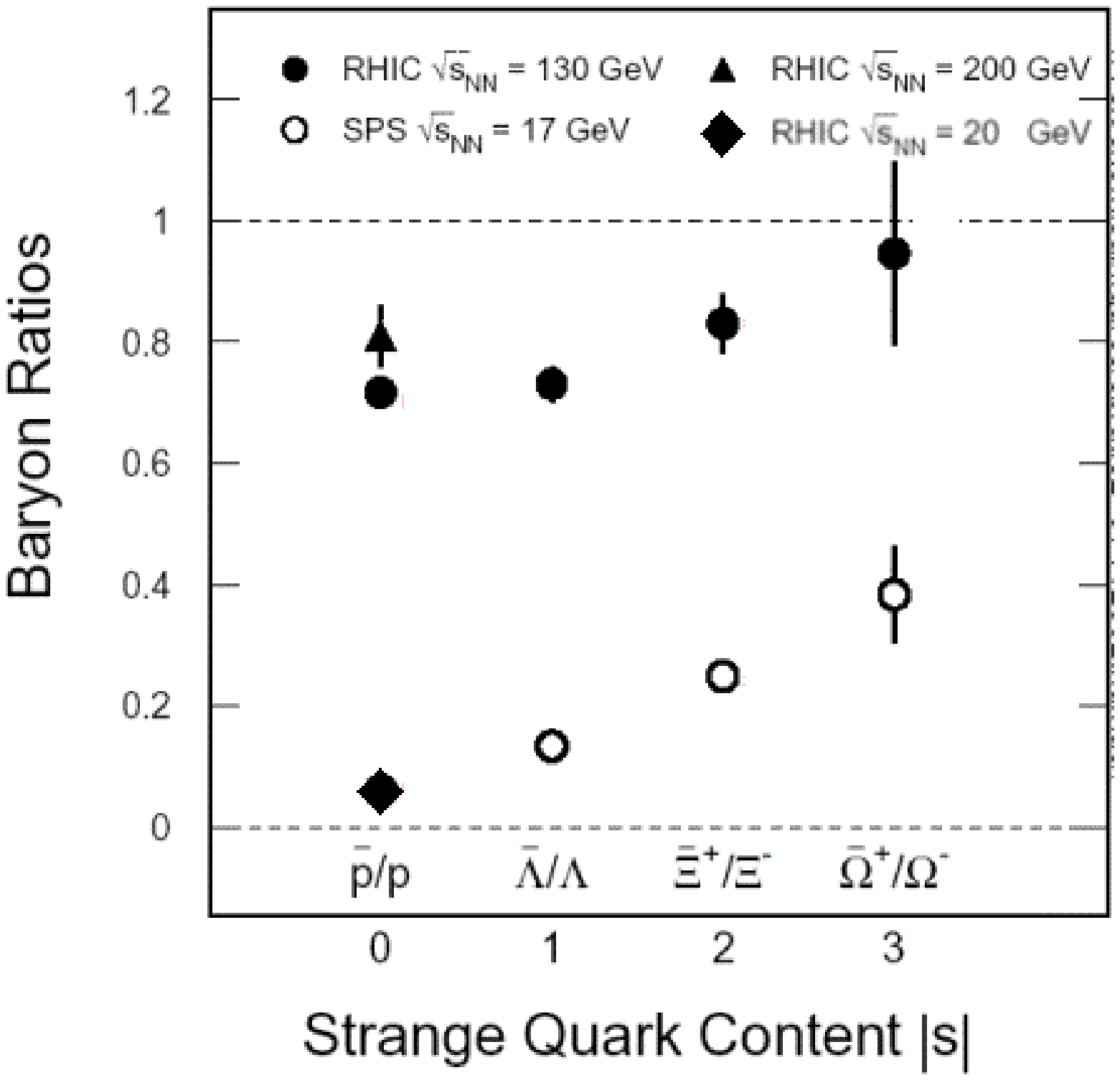}
\caption{$\bar{B}/B$ ratios versus strange quark content at RHIC (full symbols)
and  SPS (open symbols) energies.} \label{bbar}
\end{center}
\end{minipage}
\hspace{0.5cm}
\begin{minipage}{0.47\linewidth}
\includegraphics[height=\linewidth,width=\linewidth]{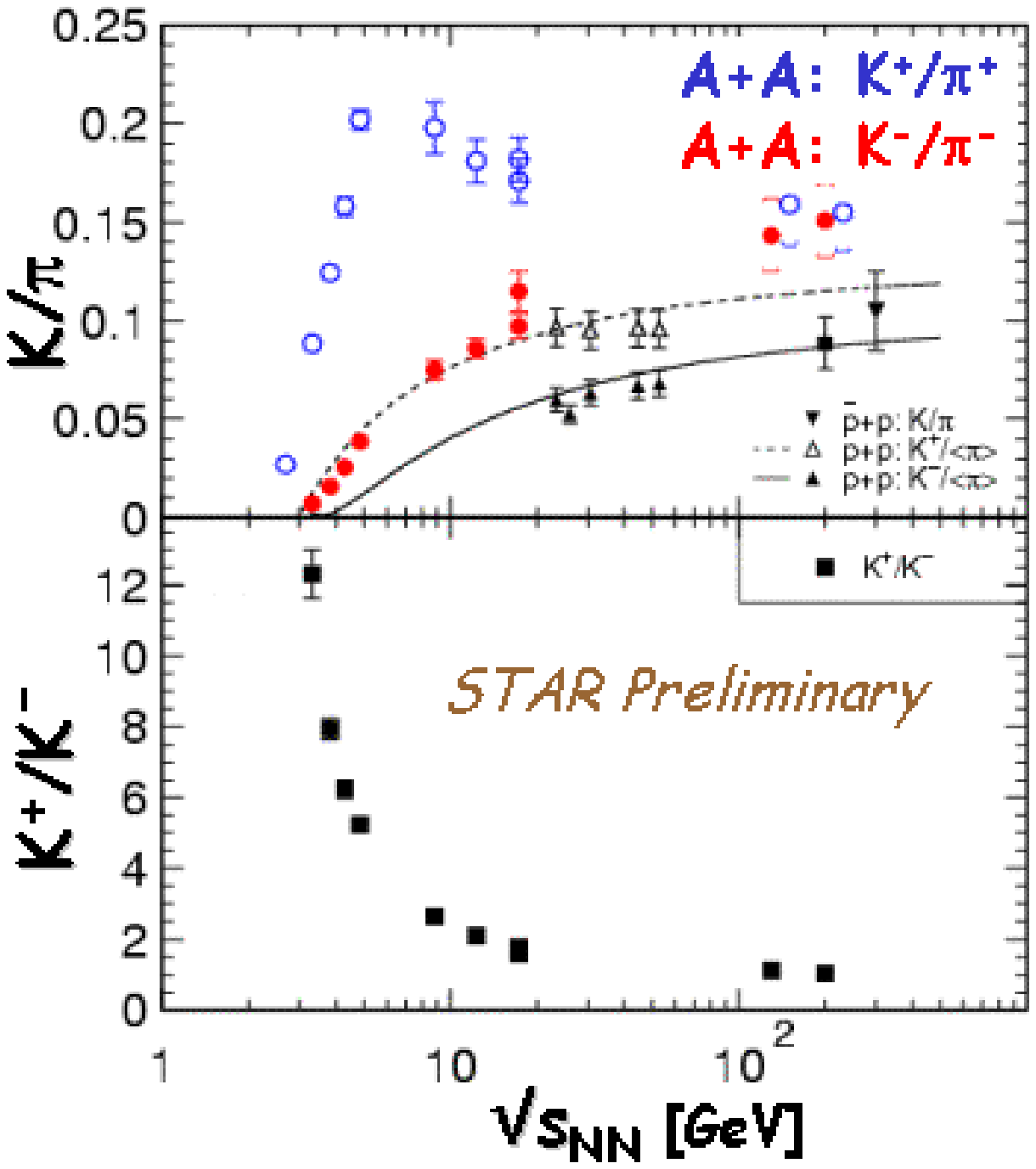}
\caption{$K$/$\pi$ (upper frame) and $K^+$/$K^-$ (lower frame) as
a function of the collision energy in the center of mass. Black
triangles represent elementary collisions.} \label{kaons}
\end{minipage}
\end{figure}

Anti-baryon over baryon ratios ($\bar{B}/B$) 
investigated as a function of the strangeness content of the
particle species are represented on figure~\ref{bbar} for
SPS~\cite{sps_ratios} and RHIC energies. Two observations can be made :
ratios increase a) with the strange content of the particle
and b) with the collision energy - due to the decreasing net baryon density. At
RHIC, the tendency clearly reveals that values tend to unity but
without reaching it. Probably, only the future LHC
collider will offer a baryon free regime.

The influence of the baryon density can also be observed on
the lower part of figure~\ref{kaons} presenting the $K^+/K^-$
excitation function sharply decreasing, as expected, with the
collision energy increase. Moreover, the $K^+/ \pi$ ratio peaking
around 8 GeV traduces the interplay between the dropping net
baryon density with $\sqrt{s_{NN}}$ and a increasing $K^+$ and
$K^-$ production rate i.e. an enhanced strangeness production. This
shape was predicted by thermal models~\cite{pbm}.

This last statement is fully responsible of the $K^-$/$\pi$ steady
increase. Note that for elementary collisions, the same trend is
observed with a lower magnitude.

\subsection{Production mechanisms}
\begin{figure}
\begin{minipage}{0.5\linewidth}
\begin{center}
\includegraphics[width=\linewidth]{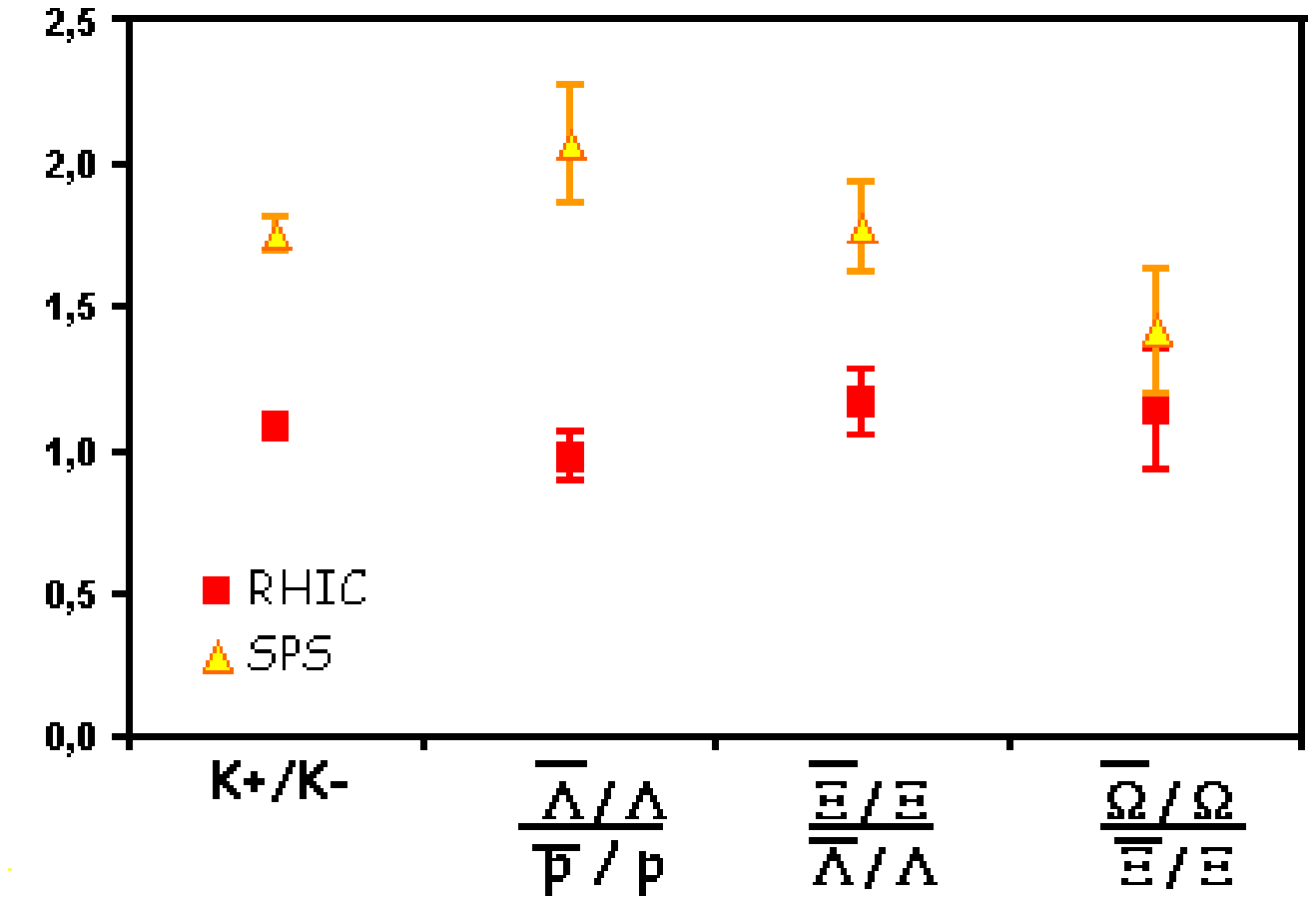}
\caption{Comparison of the measured $K^+/K^-$ ratio with those 
calculated from other particle ratios at SPS (triangles) 
and RHIC (squares) energies.} \label{ratio_alcor}
\end{center}
\end{minipage}
\hspace{0.5cm}
\begin{minipage}{0.47\linewidth}
\includegraphics[height=\linewidth,width=\linewidth]{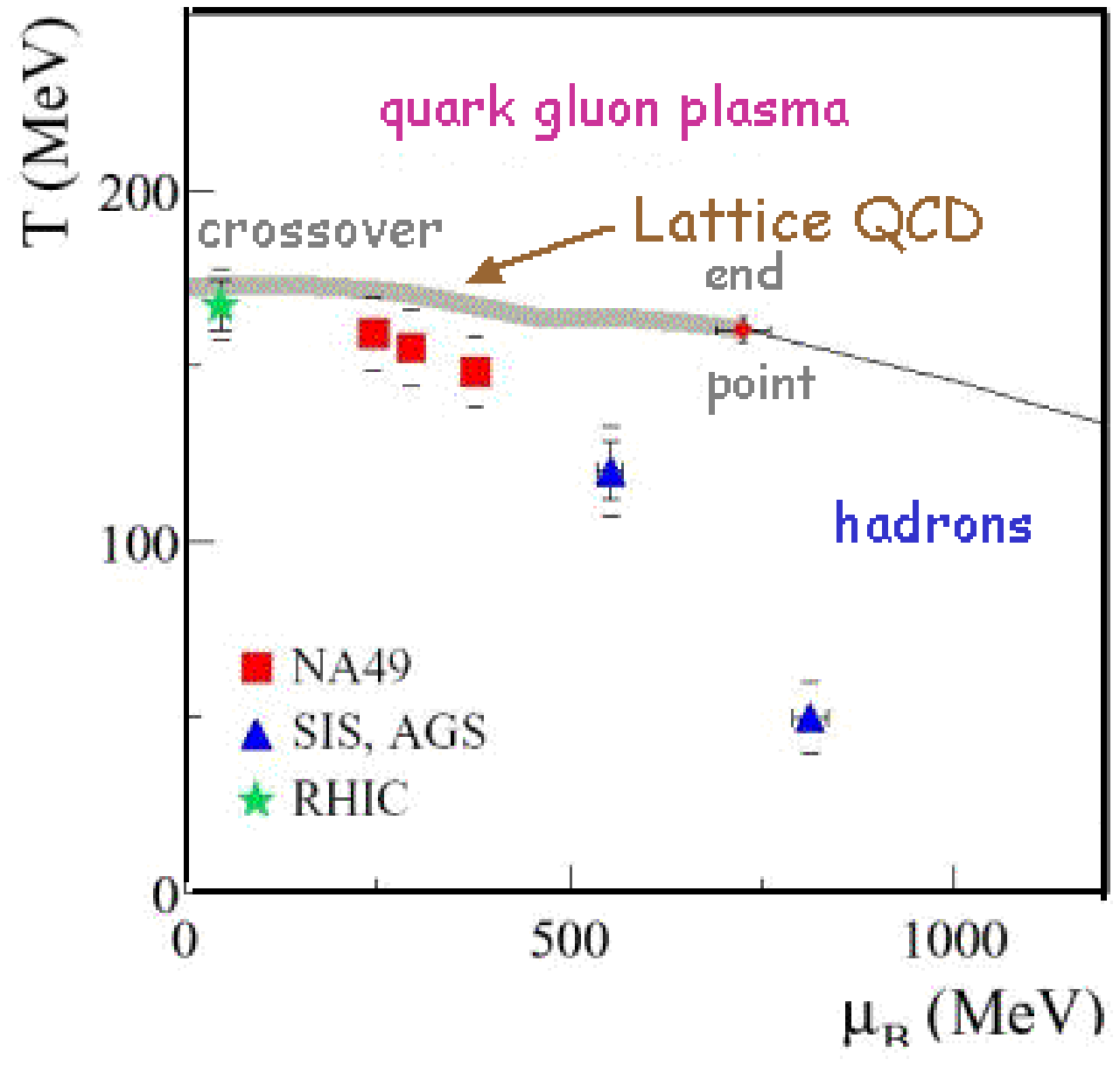}
\caption{QCD phase diagram. Chemical freeze-out parameters at
AGS, SPS and RHIC are also reported.} \label{qcd_fo}
\end{minipage}
\end{figure}

The partition between pair processes and baryon transport can be
estimated from the $\bar{p}/p$ ratio~\cite{pbarp}. Pair processes
mean pair creation as well as annihilation while baryon transport
is related to nucleons originated from the incident nucleus. It
appears that, at RHIC, pair processes are dominant, being much larger
than baryon transport by a factor of 4 while at SPS, it represents only 
20\%.

Experimental data have been compared to coalescence model
predictions~\cite{zimanyi}. Their authors claim that (anti-)quark
matter hadronizes suddenly via quark coalescence processes and that
$\bar{B}/B$ ratios are related one to each other by a multiplicative
factor $D$ given by the value of $K^+/K^-$ ratio :

$$\frac{\bar{\Lambda}}{\Lambda} = D\;\frac{\bar{p}}{p}
\quad\mbox{,}\quad \frac{\bar{\Xi}}{\Xi} = D\;
\frac{\bar{\Lambda}}{\Lambda} \quad\mbox{,}\quad
\frac{\bar{\Omega}}{\Omega} = D\; \frac{\bar{\Xi}}{\Xi}$$

These relations can be easily demonstrated by writing explicitly the
quark content of the involved particles. Hence, all ratios can be
predicted via simple quark counting. The coherence of this statement
can be seen in figure~\ref{ratio_alcor} where the measured $K^+/K^-$
ratio is found to be in line with the values obtained from other ratios,
at both SPS and RHIC energies.

An alternative of this description is provided by statistical models
cited in reference~\cite{statistical}. During the last Quark Matter
conference was discussed a compilation of P. Braun-Munzinger and
collaborators, showing various particle ratios measured at $\sqrt{s_{NN}}$
= 130 GeV by the four RHIC experiments and compared to their model
predictions. The remarkable agreement allows to extract the
temperature and baryon density at the chemical freeze-out, being 
respectively equal to 176 MeV and 41 MeV. Similar analyses have been
done at different energies using other approaches. These later certainly contain 
 some discrepancies (chemical and/or thermal equilibrium is 
supposed or not) but globally all approaches work well and thus lead
to the question whether we are dealing with chemically equilibrated
systems or not?  Superimposing the various freeze-out parameters on
the T-$\mu$ plane phase diagram of lattice QCD
calculations~\cite{qcd}, it appears that RHIC parameters coincide with
the critical value predicted by QCD as shown by
figure~\ref{qcd_fo}. For R.  Stock~\cite{stock}, {\it "This can not be
a coincidence"} while V. Koch~\cite{koch} asks the question {\it "Does
it reflect a measurement of the phase separation line in the QCD phase
diagram?"} and answers {\it "Certainly not!! Maybe it tells us about a
limiting temperature."}

Knowledge on chemical freeze-out appearing accessible,
characterization on the thermal freeze-out can be provided with
transverse mass or momentum distributions. The main result can be
summarized here, namely that the flow is found to be enhanced from SPS
to RHIC energy. The other point is that for both energies, the inverse
slope of these spectra increases linearly for light particles and
saturates for heavier masses ($>$ 1 GeV/c$^2$). Attempts of
explanation are pointed out such as the existence of a flow at a
partonic level or an earlier thermal freeze-out (decoupling) for the
heavier particles.  The measurement of other massive objects like D or
J/$\psi$ meson will be very instructive. During the conference, Matt 
Lamont has presented a review of STAR results related to the various
flow analyses~\cite{matt}.

\section{Hadronization suddenness}

\begin{figure}
\begin{minipage}{0.5\linewidth}
\begin{center}
\includegraphics[width=\linewidth]{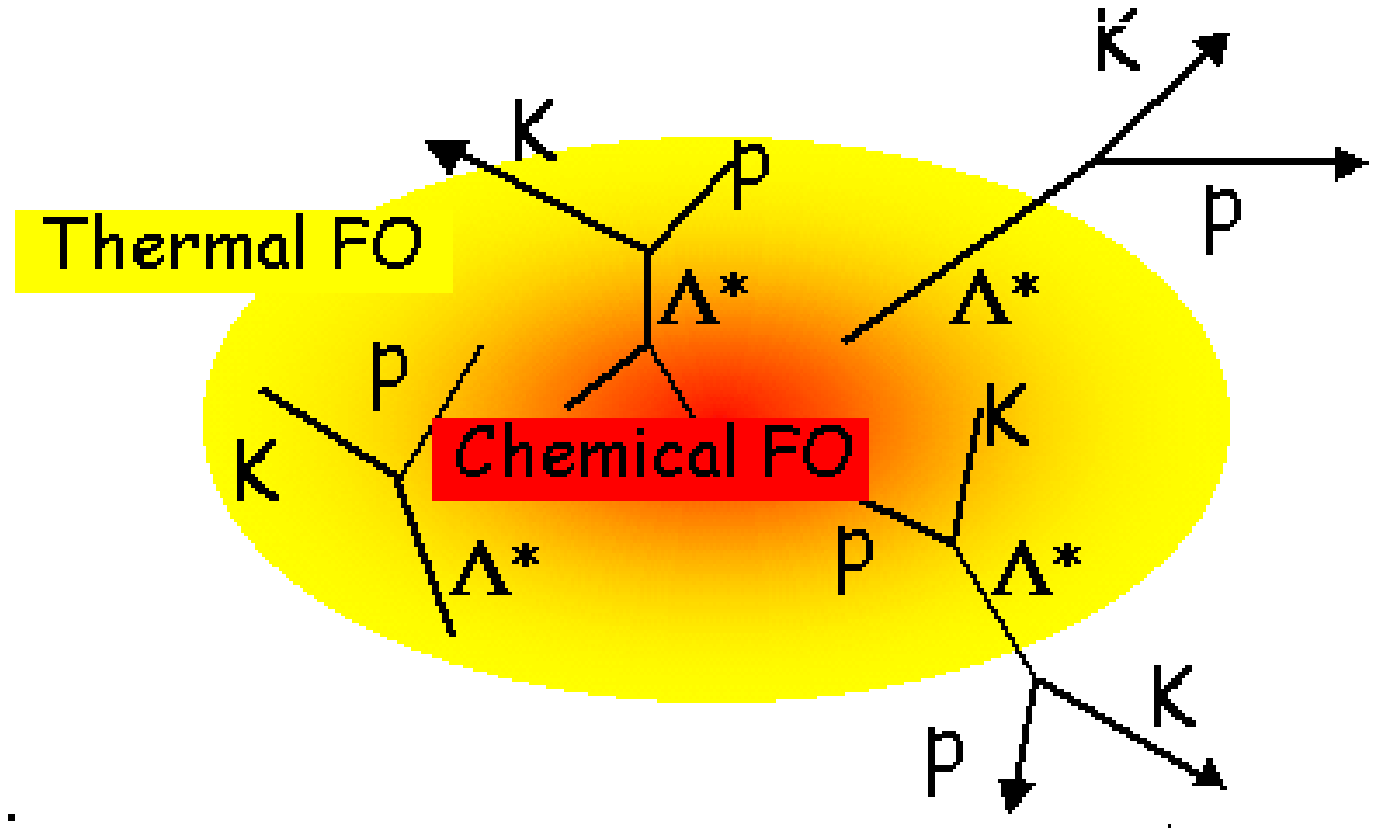}
\caption{Diagram of different cases related to eventual rescattering of
$\Lambda(1520)$ decay products.} \label{reson}
\end{center}
\end{minipage}
\hspace{0.5cm}
\begin{minipage}{0.47\linewidth}
\includegraphics[width=\linewidth]{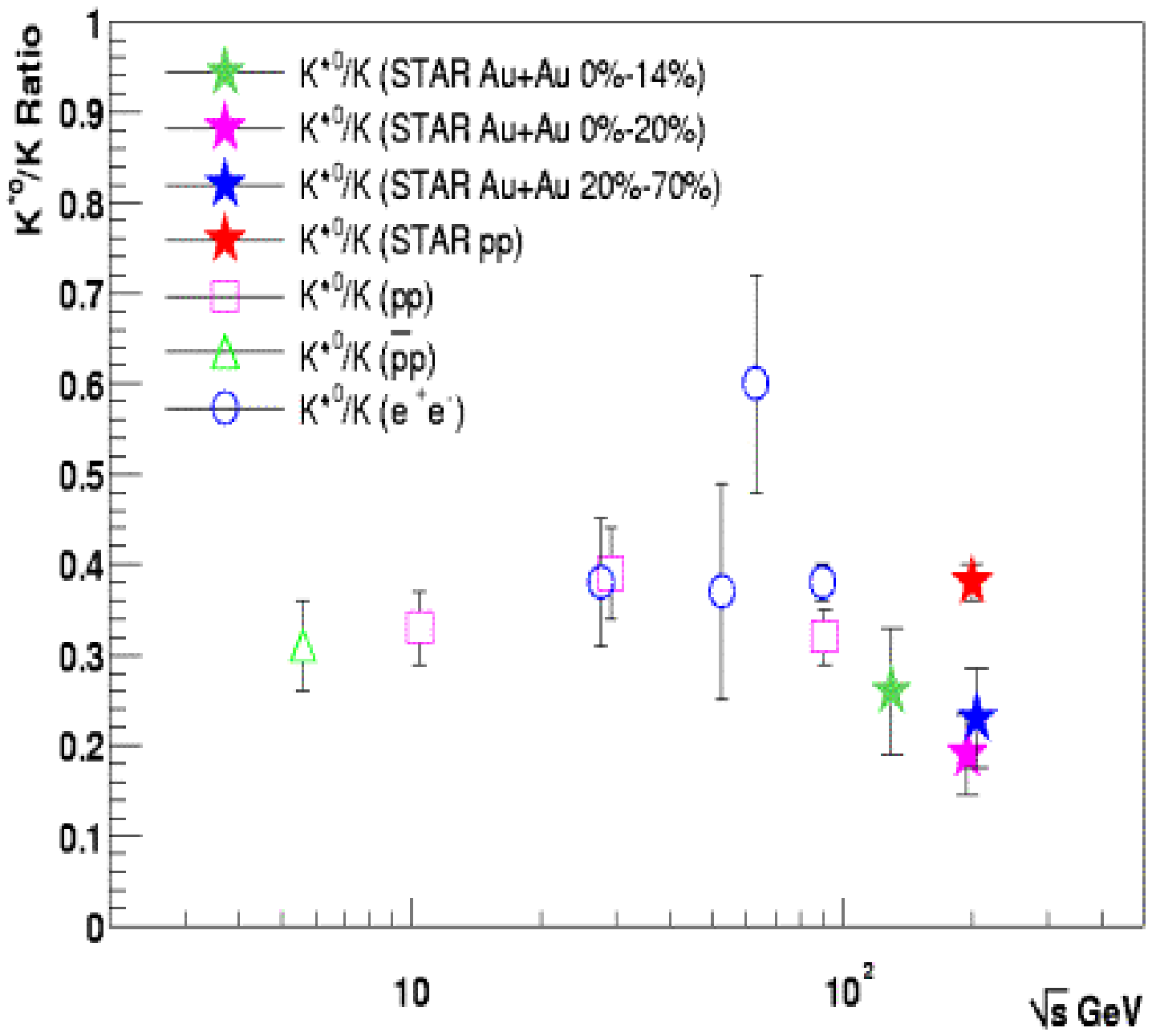}
\caption{$K^*(892)^0/K$  yield ratio as a function of colliding system
and energy.} \label{kstark}
\end{minipage}
\end{figure}

\subsection{Motivation}
Chemical and kinetic freeze-outs appear sequentially at different
times which remain to be defined. A way to investigate the duration
between both freeze-outs is provided by resonances of strange
particles as it has been suggested by J. Rafelski and
collaborators~\cite{torrieri}. The lifetime of these resonances is
similar to the lifetime of the system, typically a few fm/c. Thus,
their abundances could inform on the lifetime of the system and more
precisely on the hadronization suddenness. Indeed, resonances are
produced at the chemical freeze-out and if, after the resonance decay,
the two decay products do not suffer any rescattering, the parent
particle may be measured (for example, by invariant mass
reconstruction). However, if one or {\it a fortiori} all decay
products rescatters, it becomes impossible to reconstruct the parent
resonance. Parallel to rescattering effects, recombination of decay
particles may append and reforms a resonance which has previously
decayed. These scenarios are depicted on figure~\ref{reson}. The
longer the lifetime of the system, i.e. the longer the separation in
time between chemical and thermal freeze-outs, the more frequent these
phenomena. According to UrQMD calculations~\cite{urqmd}, the signal
loss in invariant mass reconstruction due to rescattering amounts in
AuAu collisions at $\sqrt{s_{NN}}$ = 200 GeV, 55$\%$ and 33$\%$ for
K$^*(892)^0$ and $\Lambda(1520)$ respectively.

The aim will be hence to compare the resonance yields in
AA collisions for which a separation in time is
expected to yields obtained in elementary reactions.

\vspace{1.cm}
\subsection{Sign of a resonance suppression ?}
Various resonances have been measured in STAR~\cite{fachini}. In
particular, K$^*(892)^0$ and $\Lambda(1520)$ resonances have been
investigated in pp and AuAu collisions via combinatorial of
particles for invariant mass reconstruction.

\begin{figure}
\begin{minipage}{0.5\linewidth}
\begin{center}
\includegraphics[width=\linewidth]{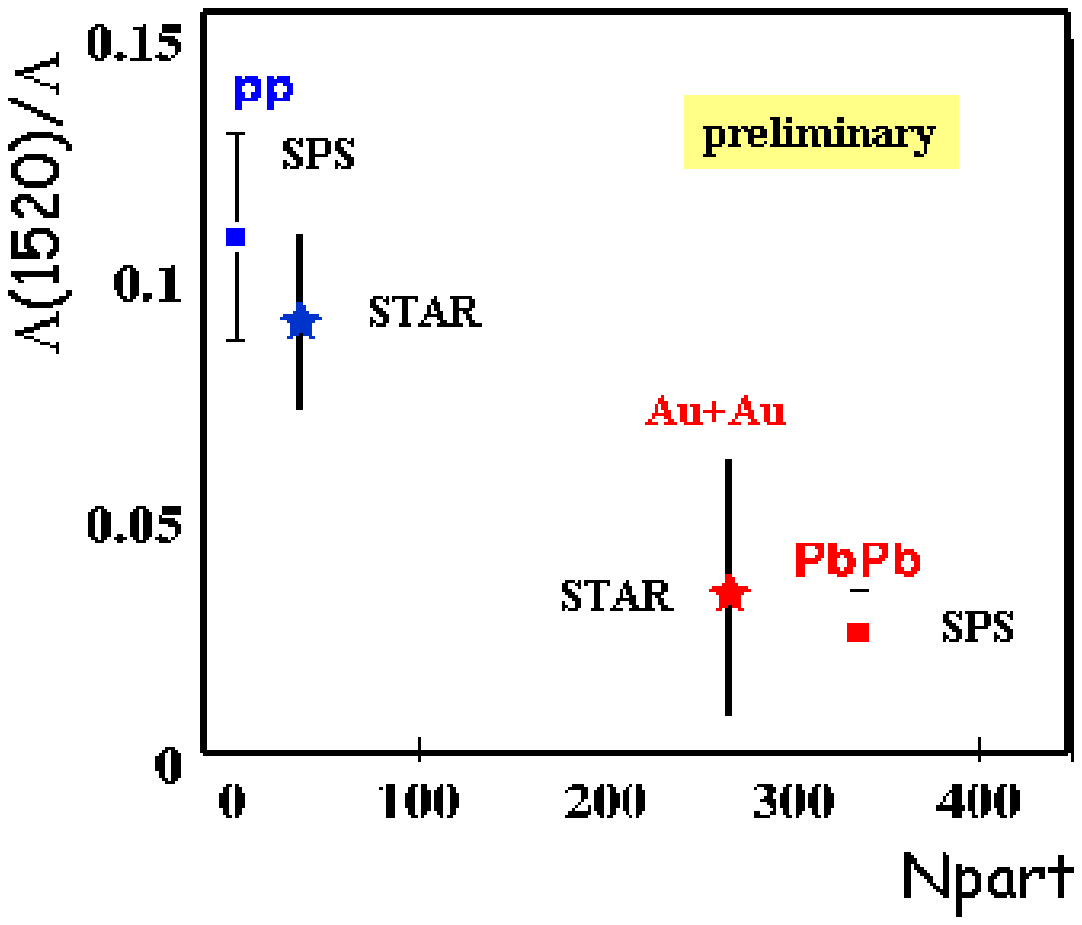}
\caption{$\Lambda(1520)/\Lambda$ as a function of the number of
participants for NA49 and STAR experiments.} \label{syst_res}
\end{center}
\end{minipage}
\hspace{0.5cm}
\begin{minipage}{0.47\linewidth}
\includegraphics[width=\linewidth]{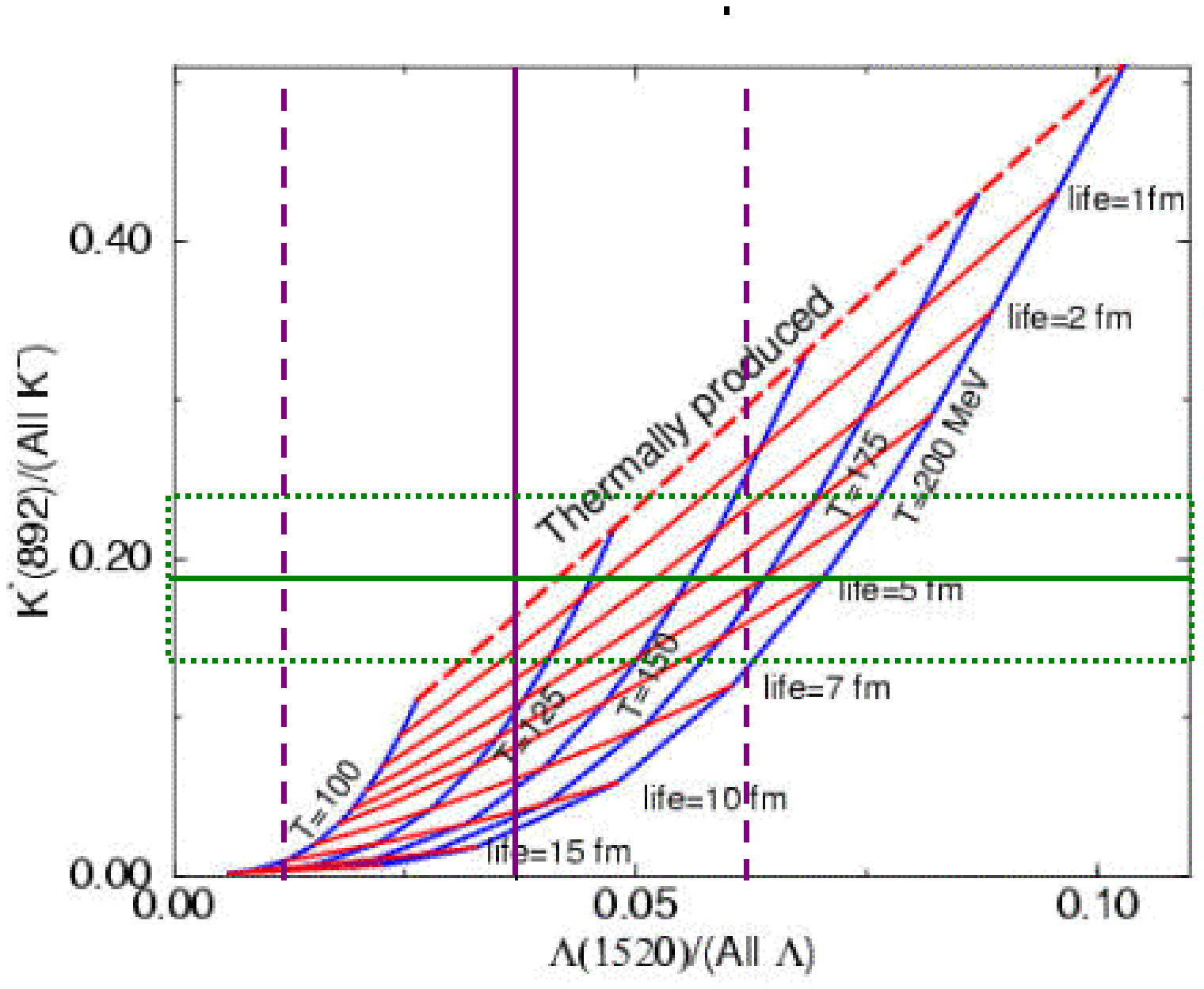}
\caption{Predicted dependence of lifetime and temperature of the system 
given by ratios K$^*(892)^0/$K and $\Lambda(1520)/\Lambda$. Straight
lines delimit STAR measurements.\vspace{1.cm}} 
\label{raf_kst_lam}
\end{minipage}
\end{figure}

K$^*(892)^0/$K ratio measurements have been studied as a function of
the beam energy and presented on figure~\ref{kstark}. The interest of
this observable resides in the fact that these particles have the same
quark content but differ by their mass and spin. K$^*(892)^0/$K ratio
appears to be lower in $\sqrt{s_{NN}}$=200 GeV AuAu collisions than in
pp at the same energy, by a factor of 2.  If one assume that this
difference is due to the K$^*(892)^0$ survival probability, this
measurement is compatible with a short time between the chemical and
thermal freeze-outs and no K$^*(892)^0$ regeneration. A long time unless 
significant K$^*(892)^0$ regeneration seems to be ruled out.

Considering now the K$^*(892)^0$ absolute yields, a suppression of
45$\%$ is observed from pp collisions (dN/dy=0.058$\pm$0.02) to the
yield scaled to the number of participants obtained for the 20$\%$
most central AuAu collisions (dN/dy=8.35$\pm$0.99). This result can be
compared to the 55$\%$ of signal loss predicted by
UrQMD. Nevertheless, two precautions have to be stressed before
concluding on the rescattering magnitude : firstly, inverse slopes of
transverse mass spectra are quite different in UrQMD and data, and
secondly, the partition between rescattering and regeneration remains
to be investigated within the model.

$\Lambda(1520)/\Lambda$ ratio has been similarly analyzed. Preliminary
results are summarized on figure~\ref{syst_res} which was recently
presented at the Breckenridge workshop~\cite{mark}. Although STAR measurements
suffer a lack of statistics, the ratio is equal to
%0.116$\pm$0.029(stat) 
0.09$\pm$0.028 in pp and 0.034$\pm$0.011(stat)
in AuAu collisions.  Thus, the signal of suppression amounts about
65$\%$ (with large error bars) being in line with UrQMD predictions as
it was already in agreement with the NA49 analysis.

Note the unexpected larger signal loss for $\Lambda(1520)$ than
for K$^*(892)^0$ whose lifetime is 3 times shorter than that of the
lambda resonances and hence should be in principle more suppressed. More precise
measurements are of course needed. This requirement appears even more crucially 
looking at figure~\ref{raf_kst_lam} where Rafelski statistical model
predictions on the lifetime and temperature of the system are
reported. K$^*(892)^0/$K and $\Lambda(1520)/\Lambda$ experimental
values are also indicated. The constraint on the model are obviously
not sufficiently severe but here, the interest of such a measurement can be
realized. 

From these resonances, one learned that there is an observed
suppression of resonances from pp to AuAu collisions and it will be
very interesting to pursue these analyses.  In particular,
$\Sigma(1385)$ (width $\Gamma_{\Sigma(1385)}$ = 35 MeV) currently under
investigation is a resonance of high interest since it is produced an
order of magnitude more abundantly than is $\Lambda(1520)$ due to an
high degeneracy factor and smaller mass. Its signal is more strongly
influenced by final state interactions than that of $\Lambda(1520)$
but not as strong as K$^*(892)^0$.

\section{Summary}
Strangeness measurement with STAR experiment has been fully
informative for the description of the hot and dense fireball. The
original environment which has been characterized corresponds to a low
baryon density but not to a baryon free regime. The various studies
related to the strange particle production indicate that coalescence
process can describe the particle creation mechanism. From the
preliminary strange resonance measurements, information on the
hadronization suddenness tends to support a very short time between
chemical and thermal freeze-outs. 

The capability of statistical models for reproducing the various
particle yields with a very nice agreement is very remarkable. It
leads to the puzzle why so ``simple'' statistical models can described
so complex relativistic heavy ion collision. STAR continues its
physics program and the new data collection related to d-Au collisions
will allow an even more complete understanding of the collision
evolution.

\end{document}